\input{aipcheck}

\documentclass[
    ,final            
  ]
  {aipproc}

\layoutstyle{8x11single}

\begin{document}

\title{Towards solving  generic  cosmological singularity problem}


\author{W{\l}odzimierz Piechocki}{address={Department of Fundamental Research, National Centre for
Nuclear Research,\\ Ho{\.z}a 69, PL-00-681 Warsaw, Poland} }

\begin{abstract}
The big bounce transition of the quantum FRW model in the setting
of loop quantum cosmology is presented. We determine the physical
self-adjoint Hamiltonian generating the dynamics. It is used to
define, via the Stone theorem, an evolution operator.  We examine
properties of expectation values of physical observables in the
process of the quantum big bounce transition. The dispersion of
observables are studied in the context of the Heisenberg
uncertainty principle. We suggest that the real nature of the
bounce may become known only after we quantize the
Belinskii-Khalatnikov-Lifshitz scenario, which concerns the
generic cosmological singularity.
\end{abstract}

\maketitle


\section{Introduction}

Observational cosmology suggests that the Universe has been
expanding for almost 14 billion years and emerged from a state
with extremely high energy densities  of matter fields.
Theoretical cosmology shows that almost all known general
relativity models of the Universe (Lema\^{i}tre, Kasner, AdS,
Friedmann, Bianchi, ..., BKL) predict an existence of cosmological
singularities  with blowing up gravitational and matter fields
invariants. The existence of the cosmological singularities in
solutions to GR means that this classical theory is {\it
incomplete}. One expects that quantization may  heal the
singularities.

In what follows we use the canonical quantization of GR based on
loop geometry. We apply the so-called {\it nonstandard} LQC, i.e.
the reduced phase space quantization approach (`first solve
constraints then quantize') in LQC, which has been developed
recently (see, e.g.
\cite{Dzierzak:2009ip,Malkiewicz:2009qv,Malkiewicz:2009zd,Mielczarek:2010rq,
Mielczarek:2011mx,Mielczarek:2012qs} and refereces therein).

\section{Reduced phase space quantization}

\subsection{Modified Hamiltonian}

The gravitational part of the Hamiltonian (in the setting of LQC)
reads
\begin{equation}\label{ham1}
    H_g:= \int_\Sigma d^3 x (N^i C_i + N^a C_a + N C) \approx 0,
\end{equation}
where $\Sigma$, space-like part of spacetime $R \times \Sigma$;
$(N^i, N^a, N)$, Lagrange multipliers; $(C_i, C_a, C)$ are Gauss,
diffeomorphism and scalar constraints; $(a,b = 1,2,3)$, spatial
indices; $(i,j,k = 1,2,3)$, internal $SU(2)$ indices.

For  flat FRW universe with massless scalar field we have
\begin{equation}\label{hamG}
H_g = - \gamma^{-2} \int_{\mathcal V} d^3 x ~N
e^{-1}\varepsilon_{ijk}
 E^{aj}E^{bk}  F^i_{ab} ,
\end{equation}
where  $\gamma$, the Barbero-Immirzi parameter; $\mathcal V\subset
\Sigma$, elementary cell; $N$, lapse function;
$\varepsilon_{ijk}$, alternating tensor; $E^a_i $, density
weighted  triad;  $ F^k_{ab} =
\partial_a A^k_b - \partial_b A^k_a + \epsilon^k_{ij} A^i_a
A^j_b$, curvature of $SU(2)$  connection $A^i_a$; $e:=\sqrt{|\det
E|}$.

Making  use of   Thiemann's identity leads finally to
\begin{equation}\label{hamR}
    H_g = \lim_{\lambda\rightarrow \,0}\; H^{(\lambda)}_g,~~~~
    H^{(\lambda)}_g = - \frac{sgn(p)}{2\pi G \gamma^3 \lambda^3}
\sum_{ijk}\,N\, \varepsilon^{ijk}\, Tr \Big(h^{(\lambda)}_i
h^{(\lambda)}_j (h^{(\lambda)}_i)^{-1} (h^{(\lambda)}_j)^{-1}
h_k^{(\lambda)}\{(h_k^{(\lambda)})^{-1},V\}\Big),
\end{equation}
where $h_k^{(\lambda)}$ is holonomy of connection around a loop
with size $\lambda$, and $V= |p|^{\frac{3}{2}}= a^3 V_0$ is the
volume of the elementary cell $\mathcal{V}$. Variables $ c$ and $
p$ determine connections $A^k_a$ and triads $E^a_k$ as follows:
$A^k_a = \,^o\omega^k_a\,c\,V_0^{-1/3} \,$ and $\,E^a_k =
\,^oe^a_k\,\sqrt{q_o}\,p\,V_0^{-2/3} $,  where $\,c = \gamma
\,\dot{a}\,V_0^{1/3}$, $\,|p| = a^2\,V_0^{2/3}$, and $\, \{c,p\} =
8\pi G \gamma /3$; scale factor is  $ a$, and  $ \dot{a}/a$,
Hubble parameter.

The total Hamiltonian for FRW universe with a massless scalar
field $\phi$ is given by
\begin{equation}\label{ham}
   H = H_g + H_\phi,~~~~H_\phi := p^2_\phi \,
|p|^{-\frac{3}{2}}/2 ,
\end{equation}
where $\phi$ and $p_\phi$ are elementary variables satisfying
$\{\phi,p_\phi\} = 1$. The relation $ H \approx 0$ enables
defining  physical phase space. Finally, the total Hamiltonian
corresponding to (\ref{ham}) reads
\begin{equation}\label{regH}
   H^{(\lambda)}/N = -\frac{3}{8\pi G \gamma^2}\;\frac{\sin^2(\lambda
\beta)}{\lambda^2}\;v + \frac{p_{\phi}^2}{2\, v},~~~~\beta :=
\frac{c}{|p|^{1/2}},~~~v := |p|^{3/2},
\end{equation}
where $\beta\sim \dot{a}/a\;$ and $\;v \sim a^3,\;$ for $\lambda =
0$. Eq (\ref{regH}) presents a  {\it modified} classical
Hamiltonian. It includes  no quantum physics.

\subsection{Dirac observables}

A function $\mathcal{O}$ defined on phase space is called the
Dirac  observable if it weakly Poisson commutes with the
constraint: $\{\mathcal{O},H^{(\lambda)}\}\approx 0$. One can show
that in the {\it physical} phase space we have only two elementary
observables satisfying the algebra: $
\{\mathcal{O}_2,\mathcal{O}_1\}= 1$.

Compound observables are functions of elementary ones. They are
supposed to be  {\it measurable} observables.  In what follows we
consider the  volume in space:
\begin{equation}\label{vvol}
v = |w|,~~~w :=
\kappa\gamma\lambda\;\mathcal{O}_1\;\cosh3\kappa(\phi-
\mathcal{O}_2).
\end{equation}
Quantization problem of $v$ reduces to the  quantization  of $w$:
\begin{equation}\label{c1}
\hat{w}\,f(x) :=
   \kappa\gamma\lambda\,\frac{1}{2}\,\big(
    \widehat{\mathcal{O}}_1\,\cosh3\kappa  (\phi-
    \widehat{\mathcal{O}}_2)
     + \cosh3\kappa  (\phi-
    \widehat{\mathcal{O}}_2)\;\widehat{\mathcal{O}}_1\big) f(x),
\end{equation}
where $f \in  L^2 (R)$. For  $\mathcal{O}_1$ and $\mathcal{O}_2$
we use the Schr\"{o}dinger representation:
\begin{equation}\label{rep1}
\mathcal{O}_1 \longrightarrow \widehat{\mathcal{O}}_1 f(x):=
-i\,\hbar\,\partial_x f(x),~~~~ \mathcal{O}_2 \longrightarrow
\widehat{\mathcal{O}}_2 f(x):= \widehat{x} f(x) := x f(x).
\end{equation}
Thus, an explicit form of $\hat{w}$ reads
\begin{equation}\label{repp1}
 \hat{w}= i\,\frac{\kappa\gamma\lambda\hbar}{2}\Big(
    2 \cosh3\kappa(\phi-x)\;\frac{d}{dx}
     -3\kappa\sinh3\kappa
    (\phi-x)\Big).
\end{equation}

Solution to the eigenvalue problem:
\begin{equation}\label{eq4}
 \hat{w}\, f_a (x) = a\,f_a (x),~~~a \in R ,
\end{equation}
is found to be \cite{Malkiewicz:2009qv}:
\begin{equation}\label{eq5}
f_a (x):= \frac{\sqrt{\frac{3\kappa}{\pi}}\exp\big(i \frac{2
a}{3\kappa^2 \gamma\lambda\hbar}\arctan
    e^{3\kappa(\phi-x)}\big)}{\cosh^{\frac{1}{2}}3\kappa(\phi-x)},
    ~~~~~~ a = b + 6\kappa^2\gamma\lambda\hbar\, m =  b + 8\pi
    G\gamma\lambda\hbar\, m,
\end{equation}
and where $b \in R,\; m\in Z$. Completion of the span of
\begin{equation}\label{set1}
\mathcal{F}_b:=\{~f_a\;|\; a = b + 8\pi G\gamma\lambda\hbar\,
m\}\subset L^2(R),
\end{equation}
in the norm of $L^2(R)$ leads to $L^2(R)$, $\forall b \in R$. The
operator $\hat{w}$ is essentially  {\it self-adjoint} on each span
of $\mathcal{F}_b$.

Due to the relation (\ref{vvol}) and the spectral theorem on
self-adjoint operators we get the solution of the eigenvalue of
the  volume operator:
\begin{equation}\label{sp1}
 v = |w|~~~\longrightarrow~~~\hat{v} f_a :=  |a| f_a .
\end{equation}
The spectrum is  {\it bounded} from below and  {\it discrete}.
There exists the minimum gap $ \bigtriangleup := 8\pi
G\gamma\hbar\,\lambda\;$ in the spectrum, which defines a {\it
quantum} of the volume. In the limit $ \lambda \rightarrow 0$,
corresponding to the   classical FRW model, there is no quantum of
the volume.

Quantization of another compound observable, the energy density
$\rho$, is presented in our paper \cite{Malkiewicz:2009qv}. The
spectrum of the corresponding operator is {\it continuous} and
{\it bounded} from above. As $\lambda \rightarrow 0$ the energy
density blows up, $\rho \rightarrow \infty$, which corresponds to
the classical case.

\section{Evolution of quantum system}
To define an {\it evolution} of the universe in a  quantum phase,
we identify first the so-called  {\it true} Hamiltonian, $H$. It
is obtain by inserting the constraint into Hamilton's equations
and finding a new (true) Hamiltonian that leads to these equations
of motion. One finds
\begin{equation}\label{trueC}
    H_\lambda =\frac{2}{\lambda\sqrt{G}} P \sin(\lambda Q),
\end{equation}
where $P :=v/(4\pi l_{{Pl}}^2 \gamma) $ and $Q := \beta$, and
where $\{Q,P\}=1$.  In the Schr\"{o}dinger representations for
these variables we have:
\begin{equation}\label{sr}
\hat{Q} \phi(Q) := Q  \phi(Q),~~~~~\hat{P} \phi (Q):= -i
\frac{d}{dQ} \phi(Q).
\end{equation}
The quantum Hamiltonian corresponding to (\ref{trueC}) reads
\begin{equation}\label{eH}
\hat{H}_{\lambda}\psi = -\frac{i}{\lambda \sqrt{G}}\left(
2\sin(\lambda Q) \frac{d}{dQ}+\lambda \cos(\lambda Q) \right)\psi,
\end{equation}
where $\psi \in D \subset \mathcal{H}:= L^2([0,\pi/\lambda],dQ)$,
and where $D$ is some  dense subspace of $\mathcal{H}$.

The eigenvalue problem, $ \hat{H}_{\lambda}\Psi_E =E \Psi_E$, has
the solution \cite{Mielczarek:2011mx}:
\begin{equation}\label{eigen}
\Psi_E(x)  =  \sqrt{ \frac{\lambda \sqrt{G}}{4\pi} \cosh\ \left(
\frac{2}{\sqrt{G}} x\right) }\;\; e^{iEx},~~~~E \in R ,
\end{equation}
where $x:= \frac{\sqrt{G}}{2} \ln \left| \tan\left( \frac{\lambda
Q}{2} \right) \right|$.

We specify the domain of $\hat{H}_{\lambda}$ as follows
\begin{equation}
\label{dom} D(\hat{H}_{\lambda}):= {\rm span} \{ \varphi_k,~~k \in
Z \},~~~~~~\varphi_k(Q):= \int_{-\infty}^\infty  c_k (E)
    \Psi_{E}(Q)\; dE,~~~c_k \in C^\infty_0 (R).
\end{equation}
One can prove that $\hat{H}_{\lambda}$ is an essentially {\it
self-adjoint} operator on $D(\hat{H}_{\lambda})$.

The classical Hamiltonian $H$ is  positive-definite because
$\lambda  \in [0,\pi]$ and $P \in [0,\infty)$. The corresponding
self-adjoint operator $\hat{H}$ has however eigenvalues $E \in R$.
We therefore introduce a  {\it physical} Hamiltonian $\hat{{H}}$,
which has only  nonnegative eigenvalues. It is defined as follows
\begin{equation}
\hat{{H}}   \Psi_{E}  := |E|   \Psi_{E},~~~~E \in R .
\end{equation}
Using Stone's theorem we define an unitary operator of the
evolution:
\begin{equation}
\hat{U}(s) = e^{- i s \hat{{H}}},
\end{equation}
where $s \in R$ is a  time parameter. The state at any moment of
time reads
\begin{equation}
  | \Psi(s) \rangle = \hat{U}(s)| \Psi(0) \rangle = e^{- i s
\hat{{H}}}| \Psi(0) \rangle .
\end{equation}

Let us consider a superposition of the Hamiltonian eigenstates $|
\Psi(0) \rangle = \int_{-\infty}^{+\infty} dE c(E) | \Psi_{E}
\rangle$ at $s=0$. Then, evolution of this state is given by
\begin{equation}\label{sup}
|\Psi(s)\rangle =\int_{-\infty}^{+\infty}  dE c(E) e^{- i s
E}|\Psi_{E}\rangle .
\end{equation}
We consider the  Gaussian packet with a simple profile defined to
be $ c(E)  := \sqrt[4]{ 2\alpha/\pi } \exp\left\{-\alpha
(E-E_0)^2\right\}$, that is centered at $E_0$ with the dispersion
parameterized by $\alpha$. The normalized packet corresponding to
(\ref{sup}) reads
\begin{equation}
\label{state} \Psi(x,s) =  \sqrt{\frac{\lambda \cosh\ \left(
\frac{2}{\sqrt{G}} x\right)} {\sqrt{8\pi \tilde{\alpha}}}}
e^{-\frac{(x-s)^2}{4\alpha}}e^{i E_0(x-s)},
\end{equation}
where $\tilde{\alpha} := \alpha/G$.

One can determine an evolution of the  {\it dispersions} $\Delta
\hat{Q}$ and $\Delta \hat{P}$, and the product $\Delta
\hat{Q}\;\Delta \hat{P}$. To see the corresponding plots, we
recommend our recent paper \cite{Mielczarek:2012qs}. It turns out
that the Heisenberg uncertainty relation, $\; \Delta
\hat{Q}\;\Delta \hat{P} \geq 1/2, \;$ is perfectly {\it satisfied}
during the entire evolution.

\section{Conclusions}

The cosmic singularity problem of  FRW model can be {\it resolved}
by using the loop geometry: big bang  turns into big bounce. The
{\it discreteness} of the spectra of the volume operator  may
favor a {\it foamy} structure of space at short distances that may
be detected in astro-cosmo observations. The {\it evolution} of a
quantum phase can be described in terms of a self-adjoint physical
Hamiltonian. The Heisenberg uncertainty relation is perfectly
satisfied during the {\it entire} evolution of the universe.

The great challenge is {\it quantization} of the
Belinskii-Khalatnikov-Lifshitz (BKL) scenario
\cite{BKL1,BKL2,BKL3}. One shows that the FRW metric is
dynamically {\it unstable} in the evolution towards  the
singularity (breaking of isotropy and homogeneity). The BKL
scenario concerns an evolution of spacetime near the cosmological
{\it singularity}.  It is a {\it general} solution to GR in the
sense that it corresponds to non-zero measure subset of all
initial conditions. It is also {\it stable} against perturbation
of initial conditions. Non-singular quantum BKL theory might be
used as a {\it realistic} model of the very early Universe.
Quantization of simple cosmological models like FRW carried out
during the last decade may be treated as warming up before meeting
this challenge.


\begin{theacknowledgments}
The author would like to thank the organizers for an inspiring
atmosphere at the Meeting.
\end{theacknowledgments}



\bibliographystyle{aipproc}   


\IfFileExists{\jobname.bbl}{}
 {\typeout{}
  \typeout{******************************************}
  \typeout{** Please run "bibtex \jobname" to optain}
  \typeout{** the bibliography and then re-run LaTeX}
  \typeout{** twice to fix the references!}
  \typeout{******************************************}
  \typeout{}
 }



\end{document}